\documentclass[twocolumn,pre,floatfix]{revtex4}
\usepackage{psfrag,epsfig,amsfonts,amssymb,amsmath,wasysym,bm}
\usepackage{dcolumn}
\usepackage{bbm,psfrag}

\usepackage[normalem]{ulem}
\usepackage{color}

\newcommand{\mini}{\mathrm{min}}
\newcommand{\maxi}{\mathrm{max}}

\newcommand{\id}{\ensuremath{\mathbbm{1}}}
\newcommand{\rhomic}{\rho_{\mathrm{mc}}}

\newcommand{\hr}{{\cal H}}

\newcommand{\tr}{\mbox{Tr}}

\newcommand{\CC}{{\mathbb C}}

\begin{document}

\title{Full expectation value statistics for randomly sampled 
pure states in high-dimensional quantum systems}

\author{Peter Reimann$^{1}$ and Jochen Gemmer$^{2}$}
\address{$^{1}$Fakult\"at f\"ur Physik, 
Universit\"at Bielefeld, 
33615 Bielefeld, Germany}
\address{$^{2}$Department of Physics, 
University of Osnabr\"uck, 
49069 Osnabr\"uck, Germany}

\begin{abstract}
We explore how the expectation values $\langle\psi |A| \psi\rangle$ of 
a largely 
arbitrary observable $A$ are distributed 
when normalized vectors $|\psi\rangle$ are randomly sampled 
from a high dimensional Hilbert space.
Our analytical results predict that the distribution exhibits a very
narrow peak of approximately Gaussian shape, while the tails
significantly deviate from a Gaussian behavior.
In the important special case that the eigenvalues of $A$ satisfy 
Wigner's semicircle law, the expectation value distribution 
for asymptotically large dimensions is explicitly obtained
in terms of a large deviation function, which exhibits
two symmetric non-analyticities akin to critical points in 
thermodynamics.
\end{abstract}

\maketitle

\section{Introduction}
\label{s1}
Consider any observable $A$ of a quantum 
mechanical model system on a Hilbert space 
$\hr$ with large but finite dimension $N$.
Then the expectation value 
$\langle\psi |A| \psi\rangle$ 
will be almost identical
for the vast majority of 
all normalized vectors 
$|\psi\rangle\in\hr$.
Equivalently, $\langle\psi |A| \psi\rangle$ 
will be very close to the microcanonical
expectation value $\tr\{\rhomic A\}$ for most 
$|\psi\rangle\in\hr$,
where $\rhomic:=\id/N$ and
$\id$ indicates the identity on $\hr$.
Similar properties are also found to
apply simultaneously for several different
observables $\{A_k\}_{k=1}^K$, 
as long as their number $K$ remains much 
smaller than the Hilbert space dimension $N$.
In particular, not only the mean value but
also the statistical fluctuations (variance)
of any given observable $A$ in the 
microcanonical ensemble will thus be imitated 
practically perfectly by nearly
any single pure state $|\psi\rangle\in\hr$.
Examples of foremost interest are 
isolated many body system
at thermal equilibrium:
If such a system is known
to be in any of those typical
pure states, then all
fluctuation phenomena at thermal 
equilibrium can actually be ascribed
to quantum fluctuations.

The quantitative derivation of those 
very general results, as well as the detailed 
discussion of their above mentioned, quite 
remarkable physical implications can be 
traced back to the Ph.D. 
Thesis by Seth Lloyd in 1988 \cite{llo88},
and are called pure state quantum
statistical mechanics therein.
Closely related variations have been 
independently rediscovered and then 
further developed under the name ``canonical typicality''
or ``concentration of measure phenomena'' 
for instance in Refs. \cite{gol06,pop06,llo06,gem09,sug07,rei07,bal08,tas16}
and references therein,
while some precursory ideas may also be
attributed, e.g., to Ref. \cite{boc59}.

At the focus of our present paper is 
the so-called full expectation value 
statistics, 
i.e., the entire probability distribution 
of expectation values 
$\langle\psi |A| \psi\rangle$, 
which arise 
when normalized vectors 
$|\psi\rangle$
are randomly sampled according to
a uniform distribution
on the unit sphere in $\hr$.
The mean value of this distribution 
is given by $\tr\{\rhomic A\}$ and 
also the variance is quantitatively 
well known \cite{llo88}.
Though never explicitly worked out so far,
it is likely that the higher moments could
in principle be determined along similar
lines, but the resulting expressions are
expected to become very involved and
therefore would be of little practical 
or conceptual use.
Accordingly, rather than going for the 
moments, we will derive here an 
alternative analytical approximation 
of the full expectation value statistics
for large Hilbert space dimensions 
$N$.

As said above, the majority of the pure states
$|\psi\rangle$ imitate the microcanonical
ensemble very well and, in particular, entail expectation 
values of $A$ very close to the thermal equilibrium 
value 
$\tr\{\rhomic A\}$.
The remaining minority of states $|\psi\rangle$
thus corresponds to all the still possible
non-equilibrium situations, and
it is natural to classify them according to 
their expectation values $\langle\psi |A| \psi\rangle$.
Especially, it seems quite interesting 
to quantify the relative measure of 
the far from equilibrium states 
along these lines.
This is the main issue of our 
present work.

\section{Setup}
\label{s2}
We start by writing the observable  (Hermitian operator) 
$A:\hr\to\hr$ 
in terms of its eigenvalues and eigenvectors as
\begin{equation}
A=\sum_{n=1}^N a_n\, |n\rangle\langle n| 
\ .
\label{1}
\end{equation}
Without loss of generality, we assume that
\begin{eqnarray}
\tr\{A\}=0
\label{2}
\end{eqnarray}
and that the eigenvalues $a_n$ are
ordered by magnitude,
\begin{eqnarray}
a_1\leq a_2\leq \ldots\leq a_N \ .
\label{3}
\end{eqnarray}
Excluding the trivial case $a_1=a_N$, Eq. (\ref{2}) 
implies that $a_1<0$ and $a_N>0$.

Next, we introduce the function $y(a)$,
which will play a key role 
in all that follows.
To begin with, we choose an arbitrary but fixed
$a\in(0,a_N)$ and define
\begin{eqnarray}
g(x) & := & \frac{1}{N}\sum_{n=1}^N \frac{1}{1+x(a-a_n)}
\ .
\label{4}
\end{eqnarray}
One readily verifies that
$g(0)=1$, $g'(0)=-a<0$, 
$g(x)\to\infty$ as $x$ approaches $x_{\maxi}(a):=1/(a_N-a)$ 
from below, 
$g(x)\to\infty$ as $x$ approaches $x_{\mini}(a):=-1/(a-a_1)$ 
from above, 
and $g''(x)>0$ for all $x\in I_a:=(x_{\mini}(a),x_{\maxi}(a))$.
These properties imply
that there must be exactly one $x\in I_a$
with $g(x)=1$.
This $x$ value is henceforth 
denoted as $y(a)$.
One thus can conclude that
$y(a)>0$, that
\begin{eqnarray}
p_n (a) & := & \frac{1}{N}\frac{1}{1+y(a)\,(a-a_n)}>0
\label{5}
\end{eqnarray}
for all $n=1,...,N$, and that
\begin{eqnarray}
\sum_{n=1}^N p_n (a)  & = & 1
\ .
\label{6}
\end{eqnarray}
Analogously, in the case $a\in(a_1,0)$ there 
exists a unique $y(a)<0$ which satisfies 
%(\ref{5})-(\ref{80}),
(\ref{5}) and (\ref{6}),
while $y(a)=0$ is the only solution
of (\ref{5}) and (\ref{6}) 
in the case $a=0$.

Altogether, $y(a)$ is thus well defined 
for any given $a\in(a_1,a_N)$,
and can be obtained as
the unique solution of the
transcendental equation
\begin{eqnarray}
\frac{1}{N}\sum_{n=1}^N \frac{1}{1+y(a)\,(a-a_n)} = 1
\label{6a}
\end{eqnarray}
with the constraints $y(a)\in I_a$
and $y(a)\not=0$ unless $a=0$.

A second main ingredient of our approach 
consists in normalized random vectors  of the form  
$|\psi\rangle=\sum_{n=1}^N c_n |n\rangle$,
where the $(c_1,...,c_N)$ are uniformly distributed 
on the unit sphere in $\CC^N$ and thus
all those $|\psi\rangle\in\hr$ are equally likely.
The probability that the expectation value 
of the observable $A$ from (\ref{1}) 
assumes some fixed value $x$
can thus be written as
\begin{eqnarray}
P(x):=\int d\mu (\psi)\ 
\delta(\langle \psi|A|\psi\rangle -x) \ ,
\label{7}
\end{eqnarray}
where the integration measure
$\mu(\psi)$ in (\ref{7}) is induced 
by the above uniform probability
distribution on the unit sphere in $\CC^N$.
According to the previous findings
in  Refs. \cite{llo88,gol06,pop06,llo06,gem09,sug07,rei07,bal08}
the distribution $P(x)$ will be
very sharply peaked for large $N$,  
hence it is natural to 
write $P(x)$ in the form
\begin{eqnarray}
P(x)=\exp\{-N\,F(x)\}\ .
\label{8}
\end{eqnarray}
This equation amounts to an 
implicit definition of the 
function  $F(x)$.
To determine its detailed
properties will be a main 
objective of our paper.

For the time being,
(\ref{8}) does not amount
to any hypothesis of how $P(x)$
``scales'' for large $N$ in the spirit of 
large deviation theory \cite{tou11}.
Rather, we take
$N$ as large but fixed and then 
consider (\ref{8}) as definition 
of $F(x)$.
In order to draw conclusions about
how $P(x)$ (and thus $F(x)$)
behaves upon variations of $N$, one 
would first have to specify how the 
observable $A$ changes with $N$, 
which is in general a quite subtle 
problem in itself.
Particularly simple special
cases will be considered 
later in Sec. \ref{s4}.

%%%%%%%%%%%%%%%%%%%%%%%%%%%%%%%%%%%%%%%%%%%%%%%%%%%%
%
\section{Main results}
\label{s3}
By means of the above ensemble of random 
vectors $|\psi\rangle$, yet another 
ensemble of random vectors 
$|\phi\rangle$ is defined via
\begin{eqnarray}
|\phi\rangle & := & \frac{R|\psi\rangle}{\sqrt{\langle\psi |R^2|\psi\rangle}}
\ ,
\label{9}
\\
R & := & \sum_{n=1}^N \sqrt{N p_n (a)} \, |n\rangle\langle n|
\ ,
\label{10}
\end{eqnarray}
where the dependence of $R$ on $a$ has been omitted.
Similarly as in (\ref{7}), we denote by
\begin{eqnarray}
P_R(x):=\int d\mu_R (\phi)\
\delta(\langle \phi|A|\phi\rangle -x) 
\label{11}
\end{eqnarray}
the probability that the expectation 
value $x$ is realized, but now for the 
ensemble of normalized random vectors 
from (\ref{9}).
Accordingly, the integration measure
$\mu_R(\phi)$ in (\ref{11}) now generically 
corresponds to some non-uniform 
probability distribution on the unit sphere in 
$\CC^N$.
Quantitatively, this ``non-uniformity'' is
captured by the following key result
of our paper:
\begin{eqnarray}
d\mu_R (\phi) & = & d\mu (\phi)\, \rho(\phi)
\ ,
\label{12}
\\
\rho(\phi) & := & 
c_R\ \langle \phi| R^{-2}|\phi\rangle^{-N}
\ ,
\label{13}
\\
c_R & := & 1/\mbox{det}[R^2] 
\ ,
\label{14}
\end{eqnarray}
where $\mu (\phi)$ is the uniform integration 
measure from above, and $\rho(\phi)$
quantifies the ``density'' or ``probability distribution''
of the $|\phi\rangle$'s on the unit sphere.
Note that $R$ from (\ref{10}) is a positive 
operator due to (\ref{5}), hence
$R^{-2}:=(R^{-1})^{2}$ in (\ref{13}) is well defined
and $c_R$ in (\ref{14}) is positive.
The derivation of this result is the 
first main achievement of our paper, but since 
the details are quite technical, it has been
postponed to the Appendix.

From (\ref{5}) and (\ref{10}) 
one can infer that
\begin{eqnarray}
\langle \phi| R^{-2}|\phi\rangle
 & = &
1+y(a)a- y(a)\langle \phi| \sum_{n=1}^N a_n |n\rangle \langle n |\phi\rangle
\nonumber
\\
& = &1+y(a)\left(a-\langle \phi|A|\phi\rangle\right)
\ ,
\label{15}
\end{eqnarray}
where Eq. (\ref{1}) was exploited in the last step.
Likewise, (\ref{14}) can be rewritten as
\begin{eqnarray}
c_R & = & \mbox{det}[1+y(a)(a-A)] 
\label{16}
\ .
\end{eqnarray}
By introducing (\ref{15})  into (\ref{11})-(\ref{13}) 
one obtains
\begin{eqnarray}
P_R(x) & = & \frac{c_R\, \int d\mu (\phi)\
\delta(\langle \phi|A|\phi\rangle -x)}{[1+y(a)(a-x)]^{N}}
\ .
\label{17}
\end{eqnarray}
The integral in (\ref{17}) can be identified 
with $P(x)$ from (\ref{7}), and with (\ref{8}) 
it follows that
\begin{eqnarray}
P_R(x) & = & c_R
\exp\{-N\,G(x)\}
\label{18}
\\
G(x) & := & F(x)+\ln[1+y(a)(a-x)]
\ .
\label{19}
\end{eqnarray}
The quantitative value of $c_R$ in (\ref{16})
may be difficult to determine, but the main 
point is that it is an $x$ independent constant.

Finally, we exploit the following 
result, whose detailed derivation has been
previously provided in Ref. \cite{rei18}.
(As expounded in \cite{rei18}, a largely 
equivalent result has also been 
obtained in Ref. \cite{mul11}, though its 
actual formulation is quite 
different.
Another related, but less rigorous
investigation has been published even
earlier in Ref. \cite{fin09}).
Namely, the overwhelming
majority of all random vectors 
$|\phi\rangle$ in (\ref{9}) entail
expectation values 
$\langle\phi|A|\phi\rangle$ 
very close to the preset value 
$a$ in (\ref{5}), provided
$a$ has been chosen so that 
\begin{eqnarray}
p_1 (a),\, p_N (a)\ll 1
\ .
\label{20}
\end{eqnarray}
Conversely, if (\ref{20}) is
violated then the random vectors in 
(\ref{9}) yield a distribution of
expectation values $\langle\phi|A|\phi\rangle$ 
without any pronounced concentration about 
some particular value.
In general, condition (\ref{20}) will be satisfied
for all $a$ values within a certain interval
around zero \cite{rei18}, whose upper and 
lower limits depend on the detailed 
spectral properties of $A$ in (\ref{1}).
More precisely, there exist two threshold
values $a_{+}\in(0,a_N)$ and $a_{-}\in(a_1,0)$ 
so that (\ref{20}) is satisfied if and only if
$a\in (a_{-},a_{+})$.
As will be seen in Sec. \ref{s4},
%(see also \cite{rei18}),
the interval $(a_{-},a_{+})$ about $a=0$
is in many cases comparable or even
almost equal to the maximally 
possible interval $(a_1,a_N)$.

If $a\in (a_{-},a_{+})$
it follows that $P_R(x)$ in 
(\ref{11}) exhibits a very narrow 
maximum around $x=a$.
Since $N$ is large this implies that
$G(x)$ in (\ref{18}) must exhibit a
minimum very close to $x=a$ and
thus $G'(a)=0$ must be
fulfilled in very good approximation.
With (\ref{19}) it follows that
\begin{eqnarray}
F'(a)-y(a)=0
\ .
\label{21}
\end{eqnarray}

Next we turn to the case $a\not\in (a_{-},a_{+})$.
As mentioned below (\ref{20}), the probability 
distribution in (\ref{18}) thus exhibits
no pronounced concentration 
about  some particular value.
Since $N$ is large, the variations
of $G(x)$ in (\ref{18}) must therefore 
be small.
As a consequence, $G'(a)=0$ and thus (\ref{21}) 
will again be satisfied in very good approximation.

So far, we tacitly considered $a$ as
arbitrary but fixed. In particular,
the operator $R$ in (\ref{10}) and 
the function $G(x)$ in (\ref{19})
in general still depend on the 
choice of $a$.
However, by observing that the
relation (\ref{21}) applies to
every given $a$ value 
within the interval $(a_1,a_N)$
we can conclude from 
(\ref{21}) that
\begin{eqnarray}
F(x)=F(0)+\int_{0}^x da\ y(a) 
\label{22}
\end{eqnarray}
for all $x\in (a_1,a_N)$,
where the value of $F(0)$ is fixed by 
the normalization of $P(x)$ in (\ref{7}).

Eq. (\ref{22}) is the second main result of our paper:
The very sharply peaked expectation value statistics 
of $A$ in (\ref{7}) is governed via 
(\ref{8}) and (\ref{22}) by the function 
$y(a)$, which is implicitly defined
as the solution of Eq. (\ref{6a}).

%%%%%%%%%%%%%%%%%%%%%%%%%%%%%%%%%%%%%%%%%%%%%%%%%%%%
%
\section{Discussion and examples}
\label{s4}
In the generic case, Eq. (\ref{6a})
cannot be solved for 
$y(a)$ in closed analytical form.
However, by Taylor-expanding 
$y(a)$ in (\ref{5}) about $a=0$
and observing that (\ref{6a}) 
identically holds for all $a$, 
one can readily determine 
$y'(0)$, $y''(0)$, $y'''(0)$,...
by comparing terms with 
equal powers of $a$.
Introducing the result into (\ref{22}), 
one obtains
\begin{eqnarray}
F(x) & = &F(0)+\frac{1}{2 m_2}x^2
-\frac{m_3}{3 m_2^3} x^3 
\nonumber
\\
& & +
\frac{2(m_3^2+m_2^3) - m_2m_4}{4m_2^5}x^4+ \ldots
\label{23}
\\
m_k & := & \frac{1}{N}\sum_{n=1}^N (a_n)^k
=\frac{\tr\{A^k\}}{N}
\ .
\label{24}
\end{eqnarray}
It follows that the probability distribution
in (\ref{8}) closely resembles a 
sharply peaked Gaussian of 
variance $m_2/N$.
However, the higher order terms in
(\ref{23}) give rise to corrections
which become more and more 
important far away from the peak,
i.e., in the very unlikely tails of the
distribution.

We recall that the results
(\ref{22}), (\ref{23}) are based on 
the approximation (\ref{21}), 
which is very good but not
exact for large but finite $N$.
For instance, (\ref{23}) yields for 
the mean value (first moment) of 
$P(x)$ in (\ref{8}) the 
approximation $m_3/\sqrt{N m_2^3}$, 
while the exact value is known to be 
zero \cite{llo88}.
In other words, our present approach
may not necessarily be optimal
if one is interested in the moments 
of $P(x)$.
Rather, the main virtue of our 
results (\ref{22}), (\ref{23}) 
is to provide insight 
about the properties of 
the distribution $P(x)$ outside 
its very narrow peak region, where very 
many moments play a notable role.

For example, by 
differentiating (\ref{6a}) with respect to 
$a$, one can 
show that $y(a)$ is a monotonically 
increasing function of $a$ within 
the domain $[0,a_N)$
(the details are explicitly
worked out in Ref. \cite{rei18}).
With (\ref{7}), (\ref{8}), and (\ref{9})
it then follows that the vast majority of
all normalized vectors $|\psi\rangle$ 
with the property 
$\langle\psi|A|\psi\rangle\geq x$
must exhibit expectation values 
$\langle\psi|A|\psi\rangle$
very close to $x$ 
for an arbitrary but
fixed $x\in[0,a_N)$,
and analogously for $x\in(a_1,0]$.

Another interesting feature arises in the
very unlikely tails of $P(x)$:
Focusing on $a>a_{+}$, one can infer 
from (\ref{5}) and the discussion below 
(\ref{20}) that $p_N (a)$ 
cannot not be small.
Exploiting (\ref{5}) once more,
it follows that $y(a)(a-a_N)=-1$ 
and hence
\begin{eqnarray}
y(a)=\frac{1}{a_N-a}
\label{25}
\end{eqnarray}
will be fulfilled in very good 
approximation for all
$a\in(a_{+},a_N)$.
With (\ref{22}) we can conclude that
\begin{eqnarray}
F(x)=F(a_+) - \ln\left(\frac{a_N-x}{a_N-a_+}\right)
\label{26}
\end{eqnarray}
and with (\ref{8}) that
\begin{eqnarray}
P(x)=P(a_+)\,\left(\frac{a_N-x}{a_N-a_+}\right)^N
\label{27}
\end{eqnarray}
for all $x\in [a_{+},a_N)$.
On the one hand, this result continuously
matches for $x\to a_N$ the obvious behavior 
$P(x)=0$ for $x > a_N$,
which readily follows from 
(\ref{1}), (\ref{3}), and (\ref{7}).
On the other hand, this result explicitly
illustrates once more the pronounced
non-Gaussian behavior of $P(x)$ 
far away from the narrow peak region.
Analogous conclusions apply in 
the domain $(a_1,a_-)$.

Recalling that $p_N (a) \ll 1$ for $a\not\in(a_{+},a_N)$
(see below (\ref{20})), it seems reasonable
to expect in view of (\ref{5}) that
the approximation
(\ref{25}) will {\em not} be fulfilled 
very well for  $a\not\in(a_{+},a_N)$,
apart from a small ``transition region''
very close to $a_+$.
Furthermore, one may surmise
that for sufficiently large $N$
(and relevant choices of $A$ as a 
function of $N$, see below),
the approximation (\ref{25}) 
becomes arbitrarily good 
and the above mentioned 
``transition region'' becomes
arbitrarily small.
As a consequence, $y(a)$ may thus be
supposed to develop a non-analyticity 
at $a=a_+$, and likewise for $a=a_-$.
In the following, 
these heuristic 
conjectures
will be worked out in more 
quantitative detail.

To begin with, we introduce the function
\begin{eqnarray}
w_N(x) := \frac{1}{N} 
\sum_{n=1}^N \delta(a_n-x)
\ ,
\label{28}
\end{eqnarray}
which is normalized to unity and 
thus may be viewed as an eigenvalue 
probability distribution.
We thus can rewrite the implicit
definition of $y(a)$ from
(\ref{6a}) as
\begin{eqnarray}
& & 
\int dx \ w_N(x)\, 
%\frac{1}{N}
\frac{1}{1+y(a)\,(a-x)}=1
\ .
\label{29}
\end{eqnarray}
In order to address the above
expectations about $y(a)$, 
we next have to specify how
$w_N(x)$ changes upon variation 
of $N$.
To this end, we focus on cases
where the eigenvalue probability 
distributions from (\ref{28}) 
approach for asymptotically 
large $N$ a well-defined limit 
$w_{\infty}(x)$ 
at least as far as the 
integral on the left hand side
of (\ref{29}) is concerned.
It is thus necessary (but not 
sufficient, see below)  
that when slightly ``smearing out'' 
the delta functions in (\ref{28}) 
then $w_N(x)$ approaches a 
reasonably well-behaving
function $w_\infty(x)$ for $N\to\infty$.
Moreover, $a_1$ and $a_N$
in (\ref{3}) are supposed 
to converge for $N\to\infty$.
Without much loss of generality, 
we specifically assume that
\begin{eqnarray}
a_N=1
\label{30}
\end{eqnarray}
for all $N$.
An analogous relation for $a_1$
will not be needed in our examples 
below, since $a_1$ will already
be fixed for any given $N$
through (\ref{2}) and 
(\ref{30}).
Note that also the thresholds $a_\pm$ 
introduced below Eq. (\ref{20}) are 
in general $N$ dependent and we tacitly 
assume that they converge for $N\to\infty$.

From a different viewpoint, all
these premises may be
considered as assumptions
about how the observable $A$ 
changes upon variations of 
$N$, see also the remarks at the end of
Sec. \ref{s2}.

%%%%%%%%%%%%%%%%%%%%%%%%%%%%%%%%%%%%%%%%%%%%%%%%%%%%
%
\subsection{Example 1}
\label{s41}
As a first example we assume that
$A$ is randomly sampled from a
Gaussian orthogonal or unitary 
ensemble (GOE or GUE) \cite{bro81},
hence its spectrum satisfies
for asymptotically large $N$ a 
so-called semicircle law.
Due to (\ref{2}) and (\ref{30}) 
this means that
\begin{eqnarray}
w_\infty(x)=\frac{2}{\pi}\sqrt{1-x^2}
\label{31}
\end{eqnarray}
for $|x|\leq 1$ and $w_\infty(x)=0$
for $|x|>1$.
As a consequence, one can show that
in the limit $N\to\infty$ the unique
solution of (\ref{29}) is
\begin{eqnarray}
y(a) & = & 4\,a\ \mbox{for $a\in[a_{-},a_{+}]$\,},
\label{32}
\\
a_{\pm} & := & \pm 1/2
\ ,
\label{33}
\end{eqnarray}
while there exists no solution for 
$a\not\in [a_{-},a_{+}]$.
These results can be verified either 
by quite tedious residue 
techniques or by quite elementary 
numerical methods.
The details seem of little 
interest and are therefore omitted.

The interpretation is as follows:
In view of (\ref{8}), (\ref{22}),
and (\ref{32}), the probability
density $P(x)$ from (\ref{7}) 
approaches for large $N$ a Gaussian 
distribution with mean zero and 
variance $1/4N$
within the domain 
$x\in [a_{-},a_{+}]=[-1/2,1/2]$.
This behavior is complemented by
(\ref{25})-(\ref{27}) for
$x\in(a_+,a_N)=(1/2,1)$
and analogous formulae for 
$x\in(a_1,a_{-})=(-1,-1/2)$.
In particular, the solutions $y(a)$ from 
(\ref{25}) and (\ref{32}) as well 
as their first derivatives coincide 
at the matching point
$a=a_{+}$, while the second 
derivatives are different,
and likewise for $a=a_{-}$;
i.e., the function $y(a)$ 
indeed develops non-analyticities
at $a=a_{\pm}$ for $N\to\infty$, 
as heuristically anticipated above
Eq. (\ref{28}).

Moreover, the existence of well 
defined limits for $y(a)$ and thus 
for $F(x)$ in (\ref{22}) when $N\to\infty$
means that $P(x)$ in
(\ref{8}) satisfies a so-called
large deviation principle 
\cite{tou11} (the limiting $F(x)$ 
being called rate function
or large deviation function
in this context).
From a different viewpoint,
the role of $F(x)$ in (\ref{8})
is reminiscent of a thermodynamic 
potential in the context of equilibrium
statistical mechanic, and the
non-analyticities of $F(x)$ at $x=a_{\pm}$,
inherited from $y(a)$ via (\ref{22}),
are then somewhat similar to critical
points in the context of 
phase transitions (see also Sec. \ref{s5}).

In turn, from the asymptotic solution
(\ref{25}) in the domain 
$(a_+,a_N)$ together with
(\ref{3})-(\ref{6}) one can infer 
that all the $p_n (a)$ are small quantities
(approaching zero for $N\to \infty$)
apart from $p_N (a)$, which converges to
a positive (non-zero) value for 
$N\to\infty$.
On the one hand, this explains why
the continuum approximation
(\ref{31}) breaks down (does not
admit solutions of (\ref{29})).
On the other hand, it  suggests to
interpret the non-analyticity
of $y(a)$ as a phase transition
similar to Bose condensation:
For $x$ values beyond $a_+$, typical 
states $|\psi\rangle$ with the
property $\langle\psi|A|\psi\rangle=x$
exhibit a ``macroscopic''
population of the eigenstate 
(or -- in case of degeneracy -- eigenspace) 
belonging to $a_N$ in 
the ``thermodynamic
limit'' $N\to\infty$.

%%%%%%%%%%%%%%%%%%%%%%%%%%%%%%%%%%%%%%%%
\begin{figure}
\epsfxsize=0.92\columnwidth
\epsfbox{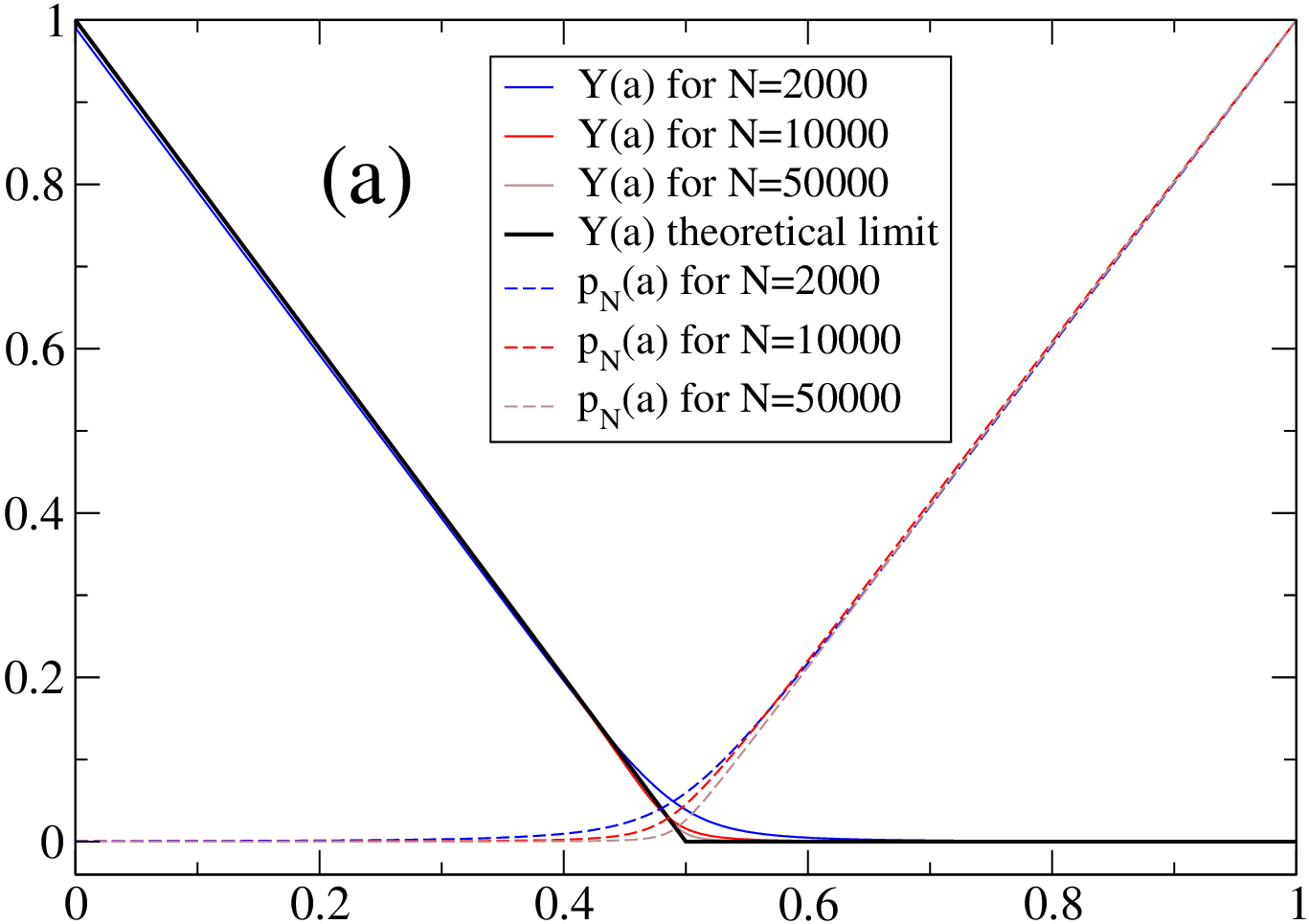}
\epsfxsize=0.95\columnwidth
\epsfbox{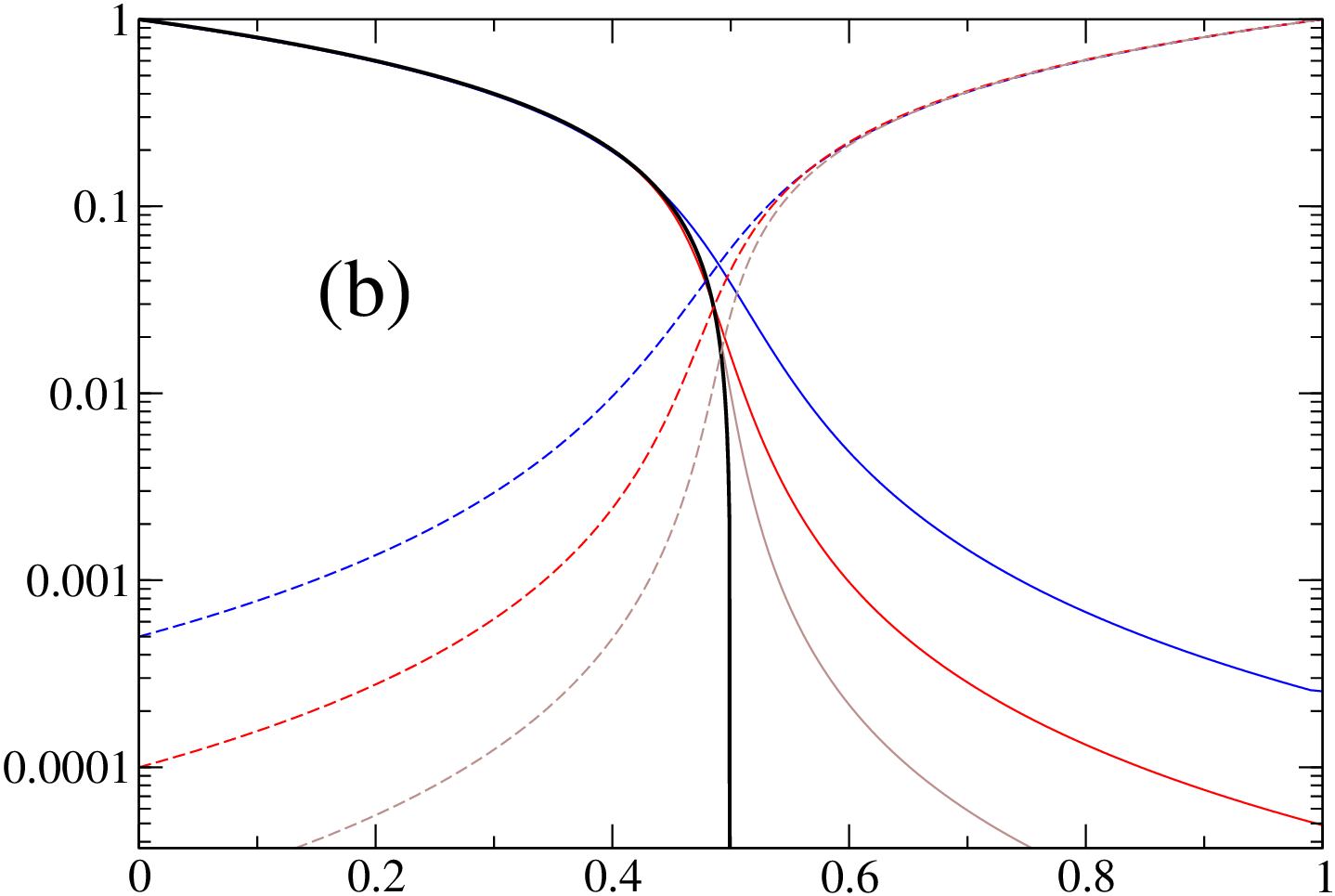}
\caption{\label{fig1}
Solid:
The function $Y(a)$ from (\ref{34}) by numerically evaluating
$y(a)$ according to 
%(\ref{5}) and (\ref{6})
(\ref{6a})
for $N=2\,000$ (blue), $N=10\,000$ (red),
and $N=50\,000$ (brown).
In each case, the $a_n$ in (\ref{6a}) 
are the eigenvalues of an $N\times N$ 
matrix, randomly sampled from a Gaussian
orthogonal ensemble (GOE) \cite{bro81}, 
and properly rescaled so that (\ref{2}) 
and (\ref{30}) are fulfilled.
Bold black line: theoretical large $N$ limit
according to (\ref{25}), (\ref{32}), (\ref{33}).
Dashed: The corresponding functions
$p_N (a)$ from (\ref{5}).
The theoretical large $N$ predictions are
$p_N (a) = 0$ for $a\in[0,1/2]$ and $p_N (1) =1$,
but are not shown in the plot.
(a) and (b): same data displayed on linear and
logarithmic scales.
The fluctuations for different samples 
of the random matrices turned out 
to be quite small (not shown).
Some remnants are still visible close to 
$a=1/2$ as apparent ``irregularities'' 
in the $N$ dependence of the curves.
}
\end{figure}
%%%%%%%%%%%%%%%%%%%%%%%%%%%%%%%%%%%%%%%%

A quantitative numerical illustration 
is provided by Fig. \ref{fig1}.
For better visibility of the details,
only $a$ values within the domain 
$[0,a_N=1]$ are shown
(the function $y(a)$ is point
symmetric about $a=0$ apart from small 
fluctuations caused by the random 
matrices).
Since the variations of $y(a)$
are unbounded (see (\ref{25}) 
and (\ref{32})) and since
the non-analyticity at 
$a_+=1/2$ is quite ``weak''
(jump in the second derivative, see above),
rather than depicting $y(a)$ itself, 
we plotted in Fig. \ref{fig1}
the quantity
\begin{eqnarray}
Y(a):=\frac{1}{4}\frac{d}{da}[y(a)\,(a_N-a)]
=
-\frac{1}{4}\frac{d}{da}
\frac{1}{N p_N (a)}
\ ,
\label{34}
\end{eqnarray}
where the last identity 
follows from (\ref{5}).
Since $y(0)=0$ (see below Eq. (\ref{6})),
the function $Y(a)$ from (\ref{34}) 
contains the same information 
as $y(a)$, but its variations
are now bounded and it makes 
the non-analyticity at $a=a_+$ 
better visible.
The (approximate) symmetry
between the solid and dashed 
curves in Fig. \ref{fig1} seems 
to be a coincidence.

%%%%%%%%%%%%%%%%%%%%%%%%%%%%%%%%%%%%%%%%%%%%%%%%%%%%
%
\subsection{Example 2}
\label{s42}
%
%%%%%%%%%%%%%%%%%%%%%%%%%%%%%%%%%%%%%%%%
\begin{figure}
\epsfxsize=0.92\columnwidth
\epsfbox{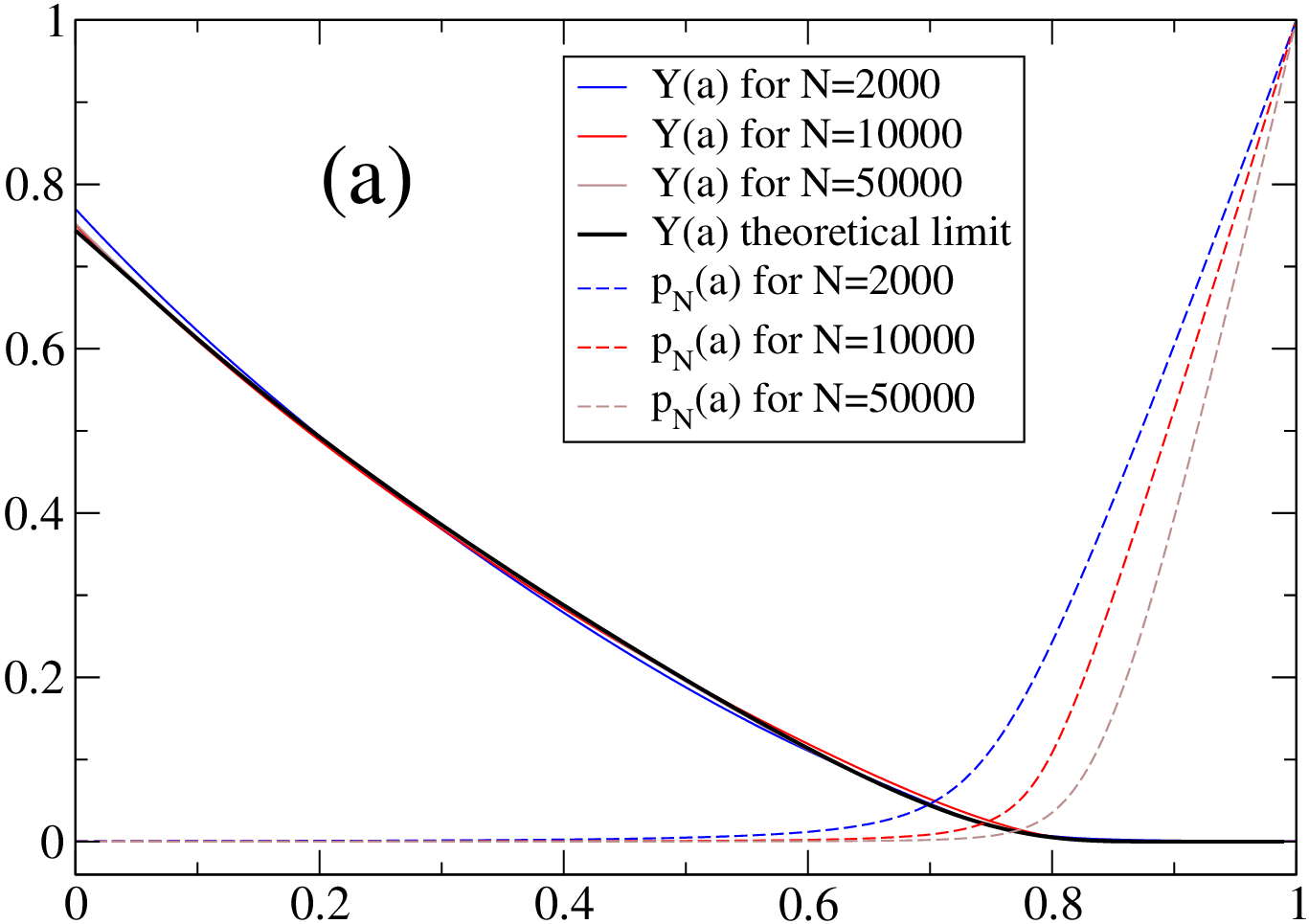}
\epsfxsize=0.95\columnwidth
\epsfbox{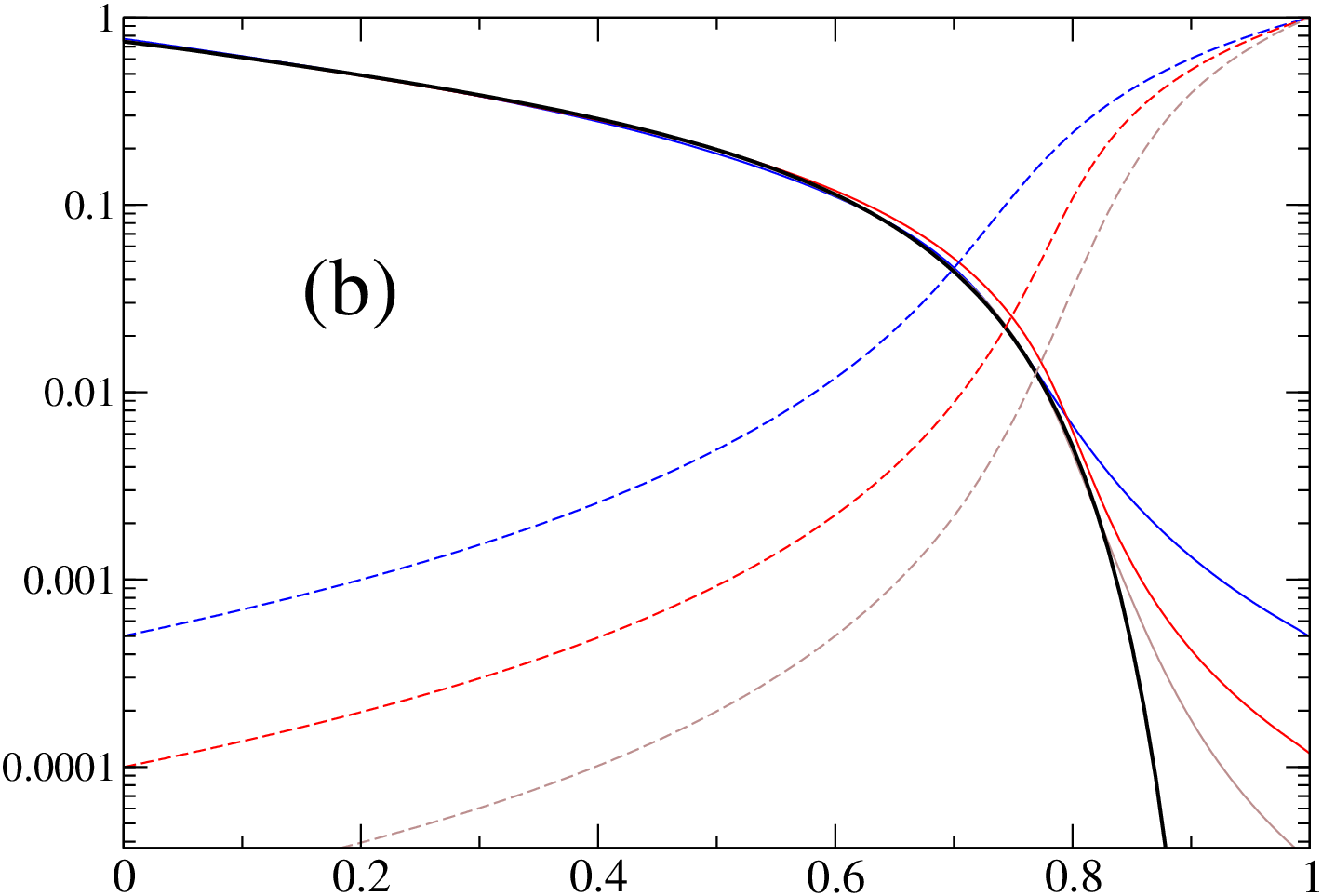}
\caption{\label{fig2}
Same as in Fig. \ref{fig1} except that the
eigenvalues $a_n$ were randomly 
generated via Wigner distributed 
differences 
$a_{n+1}-a_n$ \cite{bro81},
and that the bold line was now obtained
by numerically solving 
(\ref{36}).
Very similar results were also found
for Poisson distributed differences 
$a_{n+1}-a_n$ (not shown).
}
\end{figure}
%%%%%%%%%%%%%%%%%%%%%%%%%%%%%%%%%%%%%%%%

As a second example, we assume that the eigenvalues
of $A$ give rise to a 
%constant 
uniform
eigenvalue probability distribution.
Similarly as in (\ref{31}), this means that
\begin{eqnarray}
w_\infty(x)=1/2
\label{35}
\end{eqnarray}
for $|x|\leq 1$ and $w_\infty(x)=0$
for $|x|>1$.
The corresponding relation 
(\ref{29}) in the limit $N\to\infty$
gives rise to the following 
transcendental equation for $y(a)$:
\begin{eqnarray}
\frac{1+y(a)\, (1+a)}{1-y(a)\, (1-a)} = e^{2y(a)}
\ .
\label{36}
\end{eqnarray}
Similarly as below (\ref{4}),
the existence and uniqueness 
of a (non-trivial)
solution $y(a)$ readily 
follows for any given 
$a\in(a_1,a_N)=(-1,1)$.
Moreover, one can show 
that $y(a)$ is analytic
and monotonically increasing
within the entire domain 
$(a_1,a_N)$.
To explicitly solve (\ref{36}) for
$y(a)$ is no longer possible in 
closed analytical form,
but is straightforward by 
numerical means, see 
Fig. \ref{fig2}.

Finally, it can be shown \cite{rei18}
that $p_N(a)$ tends to zero
for $N\to\infty$ and 
any given $a\in(a_1,a_N)$.
In particular, $a_+$ approaches
$a_N=1$ for $N\to\infty$, i.e., the
situation is now reminiscent of a so
called quantum phase transition
(occurring at zero temperature).
A quantitative illustration is
depicted in Fig. \ref{fig2}.

%%%%%%%%%%%%%%%%%%%%%%%%%%%%%%%%%%%%%%%%%%%%%%%%%%%%
%
\subsection{Outlook}
\label{s43}
It seems reasonable to expect that the behavior
will be qualitatively similar to Fig. \ref{fig1}
whenever the spectrum of $A$ can be 
approximated by an eigenvalue probability
distribution $w_{\infty}(x)$ 
which approaches zero 
for $x\to a_N$.
Otherwise, a behavior similar to
Fig. \ref{fig2} is expected.
Analogous conclusion are also suggested
by Appendix D of Ref. \cite{rei18}.

More precisely, we recall that $w_{\infty}(x)$ 
approaches zero as $x\to a_N$ proportional
to $(a_N-x)^\gamma$ 
with $\gamma=1/2$ in the
example depicted in Fig. \ref{fig1}
(see also Eq. (\ref{31})).
Upon decreasing the exponent $\gamma$, 
the position of the non-analyticity 
(at $x=1/2$ for the example in Fig. \ref{fig1})
is, roughly speaking, expected to increase
until it hits the upper limit $a_N$ when
$\gamma=0$.

%%%%%%%%%%%%%%%%%%%%%%%%%%%%%%%%%%%%%%%%%%%%%%%%%%%%
%
\section{Summary and Conclusions}
\label{s5}
Given a Hermitian operator (observable) $A$
on a high-dimensional Hilbert space $\hr$,
what is the probability distribution $P(x)$ 
of the expectation values $\langle\psi |A| \psi\rangle$
when normalized vectors (pure states) 
$|\psi\rangle$ are randomly sampled according to 
a uniform distribution on the unit sphere in $\hr\,$?
The answer clearly depends on the 
spectrum of $A$ (and on nothing 
else), but general quantitative statements
are far from obvious.
The main achievement of our paper
is to show that $P(x)$ is connected to
the eigenvalues $a_1,...,a_N$ of $A$ 
via the transcendental 
equation (\ref{6a}), whose solution $y(a)$
determines $P(x)$ 
according to (\ref{8}) and (\ref{22}).

It has been previously established in Refs.
\cite{llo88,gol06,pop06,llo06,gem09,sug07,rei07,bal08,tas16}
that $P(x)$ exhibits a narrow peak
about the microcanonical expectation 
$\tr\{A\}/N$,
which can be set to zero
without loss of generality, so that the
variance of the peak is (approximately)
given by $\tr\{A^2\}/N^2$.
In other words (see also Introduction), 
the majority of states $|\psi\rangle$  behave
practically indistinguishable from
the microcanonical ensemble  and
may thus be considered as equilibrium 
states.
Accordingly, the remaining minority of 
$|\psi\rangle$'s are the non-equilibrium
states, and $P(x)$ quantifies how
many of them assume the
non-equilibrium expectation value 
$\langle\psi |A| \psi\rangle=x$.
In particular, the tails of $P(x)$ 
provide interesting information 
about the relative rareness
of the far from equilibrium states.

We have shown that $P(x)$ exhibits
an approximately Gaussian shape
within the narrow peak region about $x=0$, 
while the tails of $P(x)$ significantly deviate 
from a Gaussian behavior.
Moreover, we found that $F(x)$
in (\ref{22}) is a strictly
monotonically increasing function
of $|x|$, implying that the 
fraction of states $|\psi\rangle$
with a given non-equilibrium
expectation value
$\langle\psi |A| \psi\rangle$
decreases exponentially fast
as this expectation value moves 
away from the equilibrium value.

In general, the transcendental 
equation (\ref{6a}) cannot be explicitly
solved for $y(a)$ in closed analytical
form. 
Yet, further progress is possible
under the assumption that the 
spectrum of $A$ can be 
adequately approximated 
(as detailed in Sec. \ref{s4})
by a well defined eigenvalue distribution
function $w_\infty(x)$ for asymptotically 
large Hilbert space dimensions $N$.
Accordingly, the function $F(x)$
in (\ref{22}) then plays the role of a
so-called large deviation function.
For example, if $A$ is a typical random
matrix from the Gaussian unitary or
orthogonal ensemble then the 
distribution $w_\infty(x)$ is determined
by Wigner's semicircle law,
and $F(x)$ exhibits two symmetric 
non-analyticities, connecting
the Gaussian peak region
of $P(x)$
with the two distinctly non-Gaussian tails.
A qualitatively similar behavior is expected
whenever the observable $A$ is so that
$w_\infty(x)$ vanishes as
$x$ approaches the upper 
or the lower end of the spectrum.
Conversely, if $A$ is so that
$w_\infty(x)$ remains non-zero 
as $x$ approaches the upper 
or the lower end of the spectrum,
then non-analyticities of $F(x)$ are
not to be expected,
and likewise for the transition 
between the Gaussian peak region 
and the non-Gaussian tails of $P(x)$.

It is worth mentioning that we were
not able to establish any physically
meaningful connection between our 
present findings and the realm of 
equilibrium thermodynamics.
In particular, there does not seem to exists
a sensible relation between the
function $P(x)$ and the key quantities in
thermodynamics, namely Boltzmann's
entropy or any other thermodynamic 
potential.
Despite this dissimilarity on
the {\em physical} level, 
there are some remarkable similarities 
on a more {\em formal} level.
Namely, 
the large $N$ limit corresponds
to the thermodynamic limit,
and
the large deviation function
$F(x)$ in (\ref{8}) plays a role analogous
to that of the entropy in thermodynamics:
it quantifies the logarithm of the
state space volume (here the 
unit sphere in $\hr$) 
which exhibit some common property
(here a common expectation value).
Accordingly, the non-analyticities of $F(x)$
may be viewed as the analogues 
of critical points in 
thermodynamics.

It finally may be pointed out
once more that our results are certainly 
of particular interest for, but not at all 
restricted to
closed many-body systems.

\begin{acknowledgments}
This work was supported by the 
Deutsche Forschungsgemeinschaft (DFG)
under Grant No. RE 1344/10-1 and
within the Research Unit FOR 2692
under under Grant No. 355031190.
\end{acknowledgments}

%%%%%%%%%%%%%%%%%%%%%%%%%%%%%%%%%%%%%%%%%%%%%%%%%%%%%%%%
\appendix
\section{}
\label{app1}
In this appendix we derive the relations
(\ref{12})-(\ref{14}).
We do so by essentially starting from (\ref{9}). 
In order to facilitate geometrical considerations,
we change to the explicit representation of quantum 
states and operators exclusively by real numbers:

Let $\{|\psi_n \rangle \}$ be any orthonormal 
basis of the Hilbert space $\hr$, for instance
the eigenvectors $|n\rangle$ of $A$ from (\ref{1}).
Consider the $2N$ real numbers
$\{ Re( \langle \psi |\psi_n \rangle) , 
Im( \langle \psi |\psi_n \rangle)\}$. 
Let those numbers be the components of the 
$2N$-dimensional, real vector $\vec{\psi}$. 
Let $\vec{\phi}$ be defined by a corresponding, 
completely analogous construction. 
Then (\ref{9}) may be rewritten as 
\begin{equation}
\label{a1}
\vec{\phi}
= 
\frac{\hat{R}\vec{\psi}}{\sqrt{\vec{\psi} 
\cdot  \hat{R}^T \hat{R} \vec{\psi}}}
\ .
\end{equation}
Here $\hat{R}$ is a $2N\times 2N$ real matrix,
the components of which may be found from 
$R$ in (\ref{10}), and $\cdot$ 
denotes the standard, real Cartesian dot product. 
For the sake of generality, we do not require 
that $\hat R$ is a symmetric matrix and denote its
transposed by $\hat R^T$.

The normalization of $|\psi \rangle$ caries over 
to a normalization of $\vec{\psi}$, 
i.e.,  $\vec{\psi} \cdot \vec{\psi} =1$. 
Thus, in a Cartesian coordinate system, the 
vectors $\vec{\psi}$ lay on a $2N-1$-dimensional 
unit-hypersphere, 
and so do the vectors $\vec{\phi}$. 
However, whereas the $\vec{\psi}$ are 
(by definition, see above (\ref{7})) uniformly 
distributed on the hypersphere, the 
$\vec{\phi}$ are not. 
It is the first aim of this appendix to 
find the density of the $\vec{\phi}$ on 
the hypersphere which essentially amount 
to calculating $\rho(\phi)$ from  (\ref{12}). 
Since (\ref{a1}) maps a hypersphere onto a 
hypersphere it may be regarded as a coordinate 
transformation. 
To this coordinate transformation corresponds 
a Jacobian matrix. According to standard integral calculus of many variables, the  density $\rho(\phi)$ may 
eventually be found from the (inverse of) the 
Gramian determinant of said Jacobian matrix, cf. below, (\ref{a4}). 

In order to arrive there, we start by a 
``locally Cartesian'' parameterization of 
the surface of the above 
hypersphere formed by the $\vec{\psi}$. 
To be more explicit, consider a parameterization
$\vec{\psi}(\theta_1,...,\theta_i,....\theta_{2N-1})$ 
such that $\vec{\psi} \cdot \vec{\psi} =1$ 
holds for any choice of the $\{\theta_i\}$.  
Furthermore, we require orthonormality, 
i.e., with the notation 
$\partial_i \vec{\psi} 
= \frac{\partial \vec{\psi}}{\partial \theta_i}$, 
the following is assumed to hold:
\begin{equation}
\label{a2}
\partial_i \vec{\psi}  \cdot \partial_j \vec{\psi} 
=\delta_{ij}, \quad   \partial_i \vec{\psi}  
\cdot \vec{\psi} = 0
\ .
\end{equation}
Thus the $ \{\partial_i \vec{\psi} \}$ span a 
local tangent plain to the hypersphere.
Using the analogous notation, the Jacobian 
matrix $\hat{j}$ which corresponds  to the 
transformation (\ref{a1}) may be defined by 
specifying its column-vectors:
\begin{equation}
\label{a3}
\hat{j}
:= 
(\partial_1 \vec{\phi},....,\partial_i \vec{\phi},....,
\partial_{2N-1}\vec{\phi})
\ .
\end{equation}
Obviously, $\hat{j}$ is a $2N \times (2N-1)$ 
real matrix. From this matrix the density 
$\rho(\phi)$ may be computed as 
\begin{equation}
\label{a4}
\rho(\phi) = (\mbox{det}[ \hat{j}^T \hat{j}])^{-\frac{1}{2}}
\ .
\end{equation}

In order to calculate $\mbox{det}[ \hat{j}^T \hat{j}]$ 
(the Gramian determinant) we take a little detour. 
Consider the  $2N \times 2N$ real matrix 
$\hat{J}$, defined by its column vectors as 
\begin{equation}
\label{a5}
\hat{J}:=(\partial_1 \vec{\phi},....,
\partial_i \vec{\phi},....,
\partial_{2N-1}\vec{\phi}, \vec{\phi})
\ ,
\end{equation}
which is just  $\hat{j}$ completed by the 
``radial vector''   $\vec{\phi}$ itself. 
Since all the  $\vec{\phi}$, just like the 
$\vec{\psi}$, lay on a hypersphere, 
the $\{\partial_i \vec{\phi}\}$ also span 
a local tangent plain to the sphere, not 
necessarily orthonormal though. 
Thus, nevertheless, 
\begin{equation}
\label{a6}
\partial_i \vec{\phi} \cdot \vec{\phi}=0
\end{equation}
holds. (This may be also seen more explicitly 
below, from (\ref{a11}).) 
As a consequence, the 
matrix $\hat{J}^T \hat{J}$ may be denoted in  
the following schematic form
\begin{equation}
\hat{J}^T \hat{J} =\begin{pmatrix}
\label{a7}
  \hat{j}^T \hat{j} &  0\\
  & 0\\
  & \vdots \\
   0  0\hdots  &  1
\end{pmatrix}
\ .
\end{equation}
Hence, one readily can infer that
\begin{equation}
 \label{a8}
 \mbox{det}[ \hat{j}^T \hat{j}]=\mbox{det}[ \hat{J}^T \hat{J}]
\ .
\end{equation}

In order to find the density $\rho(\phi)$ 
according to (\ref{a4}), we now aim at finding a 
more explicit form of $\hat{J}$ that will allow 
for the computation of the determinant on the 
r.h.s. of (\ref{a8}). Computing the 
column-vectors of $\hat{J}$ (from (\ref{a5}), i.e., taking 
derivatives of (\ref{a1}))
yields, for all but the last one,
\begin{eqnarray}
\partial_i \vec{\phi}
& = &
\frac{\hat{R} \partial_i \vec{\psi}}{\left(\vec{\psi} 
\cdot    \hat{R}^T   \hat{R}  \vec{\psi}\right)^{1/2}} 
- \frac{Q}{2}
\frac{\hat{R}\vec{\psi}}{ \left(\vec{\psi} 
\cdot    \hat{R}^T   \hat{R} \vec{\psi}\right)^{3/2}}
\label{a9}
\\
Q & := &
%\left(
\partial_i \vec{\psi} 
\cdot    \hat{R}^T   \hat{R} \vec{\psi}
+ 
\vec{\psi} \cdot \hat{R}^T \hat{R} \partial_i \vec{\psi}
%\right)
\ .
 \label{a10}
\end{eqnarray}
Exploiting the properties of the dot-product, 
this may be rewritten as:
\begin{equation}
 \label{a11}
 \partial_i \vec{\phi}
= \frac{\hat{R} \partial_i \vec{\psi}}{  \sqrt{\vec{\psi} 
\cdot    \hat{R}^T   \hat{R}  \vec{\psi}}} 
- \left( \vec{\phi} 
\cdot \frac{\hat{R} \partial_i \vec{\psi}}{  \sqrt{\vec{\psi} 
\cdot    \hat{R}^T   \hat{R}  \vec{\psi}}}\right) \vec{\phi}
\ .
\end{equation}
(Note, that forming the dot-product of (\ref{a11}) 
with $\vec{\phi}$ confirms (\ref{a6}).) 
It may be seen from (\ref{a11}) that the 
first $2N-1$ column vectors of $\hat{J}$
just consist of the vectors 
$\frac{\hat{R} \partial_i \vec{\psi}}{  \sqrt{\vec{\psi} 
\cdot    \hat{R}^T   \hat{R}  \vec{\psi}}}$, 
subtracted by multiples of  $\vec{\phi}$ from 
each of them. 
However,  $\vec{\phi}$ is the last column-vector 
of $\hat{J}$. 
Since, according to basic linear algebra,
adding multiples of column-vectors to other 
column-vectors of a matrix does not change 
determinant of the latter, we may conclude:
\begin{equation}
 \label{a12}
 \mbox{det}[\hat{J}]=\mbox{det}[\hat{K}]
 \ ,
\end{equation}
where  $\hat{K}$ is defined as
\begin{eqnarray}
 \label{a13}
 \hat{K} 
:= 
\frac{ (\hat{R} \partial_1 \vec{\psi},.....,\hat{R} 
\partial_i \vec{\psi},.....,\hat{R} \partial_{2N-1}\vec{\psi},\hat{R}\vec{\psi} ) }
 {  \sqrt{\vec{\psi} 
\cdot    \hat{R}^T   \hat{R}  \vec{\psi}}}
\ .
\end{eqnarray}
Here the vectors in the numerator are again 
supposed to be the column vectors of the matrix 
$\hat{K}$ and the fraction bar notation
is meant to indicate that each column-vector 
has a prefactor given by the inverse of 
the expression in the denominator. 

To repeat: (\ref{a12}) and (\ref{a13}) 
follow from (\ref{a5}) and  (\ref{a11}). 
Due to the fact that the $\{\partial_i \vec{\psi}\}$ 
together with $ \vec{\psi}$ form a complete, 
orthonormal basis (cf. (\ref{a2})), 
we may conclude that  (\ref{a13}) 
really just implements  a specific 
representation (w.r.t. said basis) of 
\begin{equation}
 \label{a14}
 \hat{K} = \frac{ \hat{R} }
 {  \sqrt{\vec{\psi} \cdot  \hat{R}^T   \hat{R}  \vec{\psi}}}  
\end{equation}
Hence the determinant of $\hat{K}$ and thus, 
according to (\ref{a12}) the determinant 
if $\hat{J}$ is
\begin{equation}
 \label{a15}
 \mbox{det}[\hat{J}]=\mbox{det}[\hat{R}](\vec{\psi} 
 \cdot \hat{R}^T \hat{R}  \vec{\psi})^{-N}.
\end{equation}
Thus, using basic properties of determinants 
and (\ref{a4}), (\ref{a8}) we find for the 
probability density
\begin{equation}
 \label{a16}
 \rho(\phi) = \frac{(\vec{\psi} 
 \cdot \hat{R}^T   \hat{R}  \vec{\psi})^{N}}{\mbox{det}[\hat{R}]}
\ .
\end{equation}
While this is almost the final result, 
the r.h.s. of (\ref{a16}) is still 
formulated in terms of $\vec{\psi}$ 
rather than in terms of $\vec{\phi}$.
In order to convert we ``invert'' (\ref{a1}):
\begin{equation}
 \label{a17}
  \vec{\psi}= \frac{\hat{R}^{-1}
  \vec{\phi}}{  \sqrt{\vec{\phi} 
   \cdot    (\hat{R}  \hat{R}^T)^{-1} \vec{\phi}}}
\ ,
\end{equation}
where existence of $\hat R^{-1}$ is justified
in the main text below (\ref{14}).
Plugging (\ref{a18}) into (\ref{a16}), we eventually obtain
\begin{equation}
 \label{a18}
 \rho(\phi)=\frac{( \vec{\phi} 
 \cdot (\hat{R}\hat{R}^T)^{-1} \vec{\phi})^{-N}}{ \mbox{det}[\hat{R}]} .
\end{equation}
Going back to standard quantum notation 
and realizing that any eigenvalue of $R$  
appears twice in the eigenvalues of $\hat{R}$,
this reads  
\begin{equation}
 \label{a19}
  \rho(\phi)
=
\frac{\langle \phi | (R   R^\dagger)^{-1} | 
\phi \rangle ^{-N}}{\mbox{det}[R^2]}
 \ .
\end{equation}
Observing that the specific operator $R$
in (\ref{10}) is Hermitian, we finally recover
(\ref{13}), (\ref{14}).

%%%%%%%%%%%%%%%%%%%%%%%%%%%%%%%%%%%%%%%%%%%%%%%%%%%%%%%

\end{document}